\documentclass[pra,twocolumn,showpacs,preprintnumbers,amsmath,amssymb,groupedaddress]{revtex4}

\usepackage{color,ulem}
\usepackage{graphicx}
\usepackage{dcolumn}
\usepackage{bm}
\usepackage[ansinew]{inputenc}
\usepackage{epsfig}
\usepackage{units}

\begin{document}

\newcommand{\spinup}{|\uparrow\rangle}
\newcommand{\spindown}{|\downarrow\rangle}
\newcommand{\bess}[1]{{J_{#1}}}
\renewcommand{\vec}[1]{\ensuremath{\boldsymbol{#1}}}
\newcommand{\Ca}{$^{40}$Ca$^+$}
\newcommand{\Caf}{$^{43}$Ca$^+$}
\newcommand{\bra}[1]{\ensuremath{\langle #1|}}   
\newcommand{\ket}[1]{\ensuremath{|#1\rangle}}   
\newcommand{\qd}{\ket{\mbox{$\downarrow$}}}
\newcommand{\qu}{\ket{\mbox{$\uparrow$}}}
\newcommand{\oqub}{\ket{\mbox{$\Uparrow$'}}}
\newcommand{\oqu}{\ket{\mbox{$\Uparrow$}}}

\newcommand{\oquu}{\ket{\mbox{$\Uparrow \Uparrow$}}}
\newcommand{\qdd}{\ket{\mbox{$\downarrow \downarrow$}}}
\newcommand{\qud}{\ket{\mbox{$\uparrow \downarrow$}}}
\newcommand{\qdu}{\ket{\mbox{$\downarrow \uparrow$}}}
\newcommand{\quu}{\ket{\mbox{$\uparrow \uparrow$}}}
\newcommand{\oqubub}{\ket{\mbox{$\Uparrow$'$\Uparrow$'}}}
\newcommand{\todo}[1]{\marginparwidth=1.0cm \marginparsep=0.1cm\marginpar{%
    \null\vspace*{-1.1\baselineskip}%
    \rule[0pt]{1.cm}{1pt}\\
    \sffamily\tiny#1\\
    \rule[2pt]{1.cm}{1pt}}}

\title{High fidelity entanglement of \Caf \ hyperfine clock states}

\author{G. Kirchmair$^{1,2}$}
\author{J. Benhelm$^{1,2}$}
\author{F. Z{\"a}hringer$^{1,2}$}
\author{R. Gerritsma$^{1,2}$}
\author{C. F. Roos$^{1,2}$}
\email{christian.roos@uibk.ac.at}
\author{R. Blatt$^{1,2}$}

\affiliation{$^1$ Institut f\"ur Quantenoptik und Quanteninformation,
\"Osterreichische Akademie der Wissenschaften, Otto-Hittmair-Platz
1, A-6020 Innsbruck, Austria\\
$^2$ Institut f\"ur Experimentalphysik, Universit\"at Innsbruck, Technikerstr.~25, A-6020 Innsbruck, Austria}

\date{\today}

\begin{abstract}
In an experiment using the odd calcium isotope \Caf \ we combine the merits of a high fidelity entangling operation on an optical transition (optical qubit) with the long coherence times of two ``clock" states in the hyperfine ground state (hyperfine qubit) by mapping between these two qubits. For state initialization, state detection, global qubit rotations and mapping operations, errors smaller than $1\%$ are achieved whereas the entangling gate adds errors of $2.3\%$. Based on these operations, we create Bell states with a fidelity of $96.9(3)\%$ in the optical qubit and a fidelity of $96.7(3)\%$ when mapped to the hyperfine states. In the latter case, the entanglement is preserved for \unit[96(3)]{ms}, exceeding the duration of a single gate operation by three orders of magnitude.
\end{abstract}



\maketitle

In experiments with trapped ions most of the elementary building blocks for quantum information processing have been demonstrated. State initialization, long quantum information storage times, entangling gates and readout have been realized with high fidelity~\cite{Leibfried2003a,Langer2005,olmschenk2007,Benhelm:2008b,Myerson:2008a,Knill:2007}. A major challenge in current experiments is to integrate these building blocks into a single setup and make them work for a given ion species and parameter range set by the trap frequencies, the magnetic field strength and further parameters. Quantum information encoded in ground state atomic levels whose energy difference only weakly depends on changes of the magnetic field ("clock" states) have been stored for more than a second~\cite{Langer2005,olmschenk2007,Benhelm:2008a,Lucas:2007}. However, high fidelity entangling operations have so far only been demonstrated on qubits with limited coherence times~\cite{Leibfried2003a,Benhelm:2008b}. Haljan \textit{et al.} \cite{Haljan:2005b} used a M{\o}lmer-S{\o}rensen entangling operation~\cite{Soerensen1999,Solano1999,Sackett2000} acting directly on two $^{111}$Cd$^+$ ions where the qubit was encoded in the $S_{1/2}(F=0,m_F=0)$ and $S_{1/2}(F=1,m_F=0)$ hyperfine ground states but the target state fidelity was limited to an average of $79\%$ by technical instabilities and laser power. An attractive way of combining high fidelity entangling gates with long storage times is to coherently map the quantum information from clock states to atomic states which are suitable for performing the entangling operation with high fidelity, and to map the information back at the end of the entangling operation~\cite{Langer2005}. In case this mapping operation is addressed at individual ions - for instance by using a strongly focused laser - this comes with the advantage that the entanglement operation can be applied to the entire ion string.

In a recent experiment we implemented a M{\o}lmer-S{\o}rensen gate operation on an optical qubit encoded in the $S_{1/2}$ ground and $D_{5/2}$ metastable state of \Ca \ and demonstrated the creation of Bell states with a fidelity of $99.3\%$~\cite{Benhelm:2008b,Kirchmair2008}. The coherence time of this optical qubit is limited by magnetic field fluctuations, laser linewidth and ultimately the 1.2~s lifetime of the $D_{5/2}$ state. For the most abundant isotope \Ca \, encoding the qubit in long-lived hyperfine clock states is impossible as this isotope has a nuclear spin $I=0$. In the isotope \Caf \ on the other hand where $I=7/2$, coherence times of many seconds have been measured for qubits encoded in the $S_{1/2}(F=4,m_F=0)$ and $S_{1/2}(F=3,m_F=0)$ states~\cite{Benhelm:2008a,Lucas:2007}.

In this paper we investigate how to combine the long quantum information storage times observed for the hyperfine qubit of a single \Caf \ ion with the high fidelity gate operation on optical qubits by mapping between these two qubits. We first give a short description of the measurement apparatus and discuss the steps for state initialization and Bell state preparation in the optical qubit. The mapping to the hyperfine qubit is described and for both qubits the entanglement decay is measured. We demonstrate remapping to the optical qubit after a 20~ms storage time, and the application of further gate operations. Due to the nuclear spin of $I=7/2$, the isotope \Caf \ has a relatively complex energy level structure compared to other ions used for quantum information processing. We discuss magnetic field and polarization settings to reduce errors due to spectator states, which are especially harmful for the gate operation in case when large coupling strengths are required in a relatively short time.

\begin{figure}[t]
\includegraphics[width=80mm]{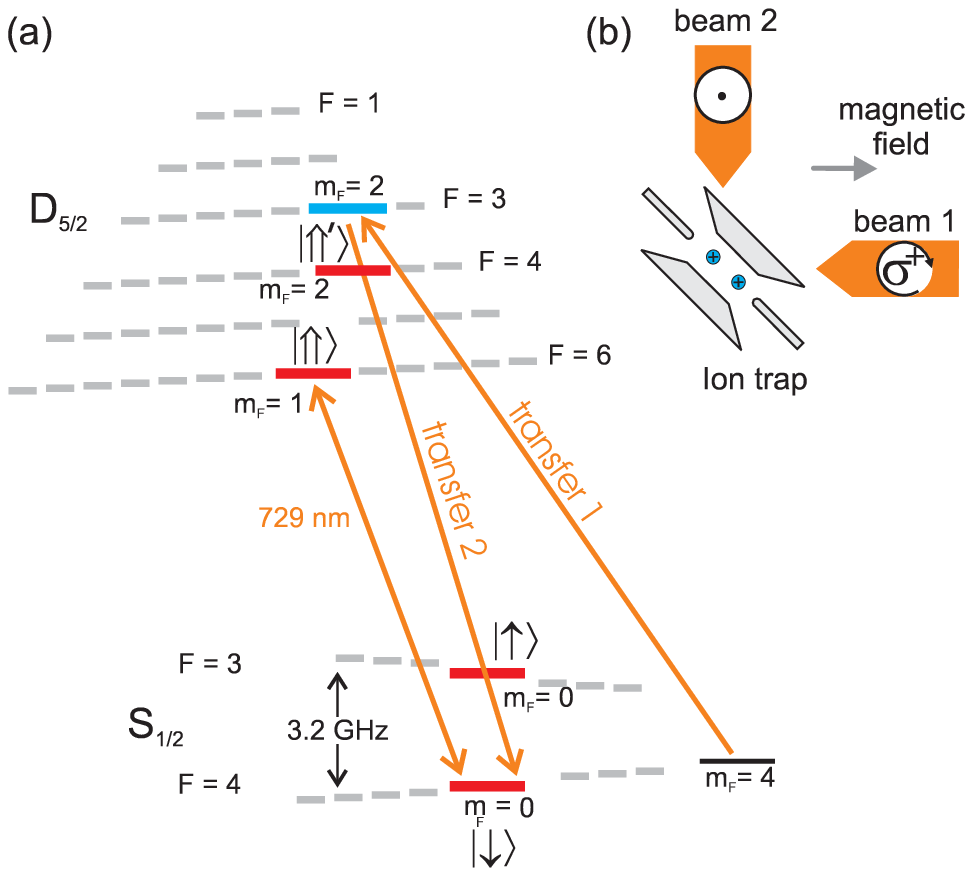}
\caption{\label{fig:levelscheme}  (Color online) (a) Energy levels of \Caf \ showing the hyperfine
splitting of the atomic states $S_{1/2}$ and $D_{5/2}$. Microwave radiation applied
to an electrode close to the ions drives the hyperfine qubit encoded in the states \qd \ and \qu. A laser at \unit[729]{nm} excites the ions on
the transition from the $S_{1/2}(F=4)$ to the $D_{5/2}(F=2,..,6)$-states. It is used for ground state cooling, state
initialization, state discrimination and to excite the optical qubit comprised of the states \qd \ and \oqu \ or \oqub .  (b) Alignment of the trap axis with respect to the magnetic field. Laser light at \unit[729]{nm} can be directed to the ions from two alternative directions with $\sigma^+$ polarization (beam~1, \oqu ) and linear polarization perpendicular to the magnetic field (beam~2, \oqub ).}
\end{figure}

A pair of \Caf \ ions is loaded from an enriched source into a linear Paul trap having a $\unit[25.5]{MHz}$ drive frequency and trapping frequencies of $(\omega_{\rm{ax}},\omega_{\rm{rad}})/2\pi\approx\unit[(1.2,2.8)]{MHz}$ in the axial and radial directions, respectively. References~\cite{Benhelm:2008a, BenhelmPhd} provide a detailed description of the ion trap apparatus and the experimental procedure was given.  A simplified energy level scheme for \Caf \ is shown in Fig.~\ref{fig:levelscheme}~(a). In the experiments reported here, the hyperfine qubit, comprised of the states $\qu\equiv S_{1/2}(F=3,m_F=0)$ and $\qd\equiv S_{1/2}(F=4,m_F=0)$, is driven by applying a microwave field ($\unit[3.2]{GHz}$) to one of the electrodes which is also used to compensate for external electric stray fields. The quadrupole transition $S_{1/2}(F=4)\leftrightarrow D_{5/2}(F=2,..,6)$ is driven with a narrow bandwidth laser whose frequency is referenced to the ions by automated service measurements occurring every 60 to \unit[120]{s} probing two different Zeeman transitions with a Ramsey type scheme. From these measurements we also infer the magnetic field strength at the position of the ions at the time of the measurement. State detection is achieved by means of a photomultiplier tube (PMT) that provides us with the three probability values $p_k$ corresponding to cases of $k$ ions emitting fluorescence when excited on the dipole transitions coupling to the $P_{1/2}$ state. The detection time of \unit[3]{ms} implies a systematic state assignment error due to decay from the $D_{5/2}$ states during detection of about $0.2\%$.

Each experimental sequence starts by Doppler-cooling the ions for \unit[3]{ms} followed by sideband cooling on the $S_{1/2}(F=4,m_F=4)\leftrightarrow D_{5/2}(F=6,m_F=6)$ quadrupole transition of both axial modes close to the ground state ($\bar{n}_{\rm com},\bar{n}_{\rm st}\approx 0.03(3), 0.4(1)$; \unit[5]{ms} and \unit[3]{ms}). An optical pumping scheme that makes use of the $D_{5/2}$ state~\cite{Benhelm:2008a} initializes the ions into the energy level $S_{1/2}(F=4,m_F=4)$ in more than $98.9\%$ of the cases.

Subsequently the populations of both ions are transferred by a $\pi$-pulse ($\unit[20]{\mu s}$) to the energy level $D_{5/2}(F=3,m_F=2)$ (transfer~1). By turning on the lasers at \unit[397]{nm} and \unit[866]{nm} \cite{Benhelm:2008a} for $\unit[500]{\mu s}$, we detect residual population remaining in the $S_{1/2}$ state manifold (PMT~check) . All cases where more than $3$ photons are detected  are rejected (we receive $\sim$~\unit[20]{photons/ms} per ion in the $S_{1/2}$-state at a dark count rate of 0.2 photons per ms). The rejection rate is typically $2\%$. Initialization of the hyperfine qubit is completed by another $\pi$-pulse (transfer~2) on the transition $D_{5/2}(F=3,m_F=2)\leftrightarrow S_{1/2}(F=4,m_F=0)\equiv \qd$. In total the initialization of internal and external degrees of freedom takes about \unit[12]{ms} and the qubit state \qdd \ is populated with a fidelity of more than $99.2\%$. The fidelity of state initialization to \qdd \ is measured by transferring the population with two consecutive $\pi$-pulses from \qdd \ to two different Zeeman-levels in the $D_{5/2}$ manifold ($D_{5/2}(F=3,m_F=2)$ and $D_{5/2}(F=5,m_F=2)$) and determining remaining population in the $S_{1/2}$ by fluorescence detection. The fidelity of optical pumping is determined in a similar way.

The magnetic field is set to \unit[6.0]{G} in order to achieve a sufficiently large frequency separation of the neighboring transition lines in the spectrum of the $S_{1/2}\leftrightarrow D_{5/2}$ quadrupole transition \cite{Benhelm2007}. The transition $\qd \leftrightarrow D_{5/2}(F=6,m_F=1)\equiv \oqu$ is chosen as optical qubit because of its small magnetic field sensitivity of \unit[350]{kHz/G} (which is at least a factor of two lower than the sensitivity of any $m_F=0$ to $m_F=0$ transition for the chosen magnetic field) and in order to avoid having coinciding resonances with (micro)motional sidebands of other spectral components. Moreover, this choice of optical qubit has the advantage of a comparatively large Clebsch-Gordan coefficient which together with a carefully set laser polarization (for this qubit we use beam~1 as shown in Fig.~\ref{fig:levelscheme}) suppresses neighboring Zeeman transitions with $\Delta m \neq +1$. The nearest transitions $\Delta m = 0$ and 2 are suppressed by factors of 38 or more in terms of their Rabi frequencies, the strongest being the $\qd \leftrightarrow D_{5/2}(F=6,m_F=0)$ transition, whose resonance is more than 4~MHz away from the $\qd \leftrightarrow \oqu$ transition. The two closest transitions are $S_{1/2}(F=4,m_F= \pm 1) \leftrightarrow \oqu$ ($\gtrsim~2$~MHz away) which are suppressed by a factor of 270 in coupling strength. The high coupling strength on the gate transition reduces the required power for the gate operation and thus the AC-Stark shift caused by off-resonant coupling to dipole transitions. This is necessary to get a high fidelity gate operation [13].

After initialization to \qdd \, a M{\o}lmer-S{\o}rensen entangling operation (MS~1) consisting of a single bichromatic laser pulse is applied to the optical qubit transition to create a Bell state of $ \qdd +i \oquu$. The gate time $\tau_{\rm gate}=\unit[100]{\mu s}$ corresponds to a detuning of \unit[10]{kHz} from the axial sideband. For the chosen beam geometry, these settings lead to an AC-Stark shift of \unit[3.5]{kHz} which is fully compensated by using different coupling strengths $\Omega_b$, $\Omega_r$ for the red and the blue detuned frequency component with $\Omega_b/\Omega_r=1.33$~\cite{Kirchmair2008}.

The fidelity of the created Bell state is determined by measuring the populations $p_0+p_2$ in \qdd \ and \oquu \ and the coherence between the two states. The coherence is inferred from the amplitude of parity oscillations that is induced by scanning the phase of an additional $\pi/2$-pulse on the optical qubit transition applied to both ions prior to detection. As a maximum fidelity, we obtain $96.9(3)\%$. The durations and errors of all experimental steps and the achieved state fidelities are given in Table~\ref{tab:budget}.

\begin{table}
\caption{\label{tab:budget} Duration and errors of each experimental step and the achieved state fidelities.}
\begin{ruledtabular}
\begin{tabular}{c|c|c|c}
\textbf{Experiment}  & \textbf{Duration}   & \textbf{Error  }            & \textbf{State }    \\
\textbf{step}   & \textbf{($\mu s$)}    & \textbf{(\%)}              & \textbf{fidelity (\%)}   \\ \hline\hline
Opt. pumping    & $600$               & $< 1.1$ & $ > 98.9 $      \\ \hline
Transfer 1      & $20$               & $0.7$ & $ 98.2 $      \\ \hline
PMT check       & $500$               & $0.7$ & $ 99.3 $   \\ \hline
Transfer 2      & $20$             & $< 0.1$ & $ > 99.2 $  \\ \hline
Total prep. \qd & $1140$              & $< 0.8$ & $ > 99.2 $   \\ \hline
MS 1            & $100$               & $2.3$ & $ 96.9 $   \\ \hline
Map             & $120$               & $0.2$ & $ 96.7 $   \\ \hline
Wait            & $20\times 10^3$     & $2.2$ & $ 94.6 $   \\ \hline
Map$^{-1}$      & $120$               & $0.7$ & $ 93.9 $   \\ \hline
MS 2            & $100$               & $3.7$ & $ 90.4 $   \\ \hline
MS 3            & $100$               & $2.9$ & $ 87.8 $
\end{tabular}
\end{ruledtabular}
\end{table}

\begin{figure}[t]
\includegraphics[width=80mm]{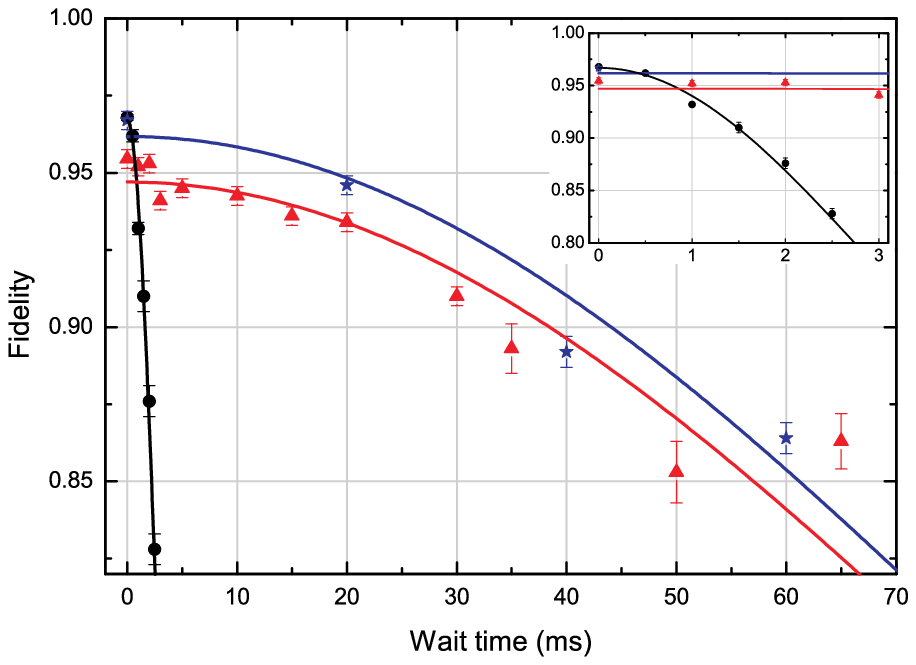}
\caption{\label{fig:decay}  (Color online) Using a bichromatic laser beam, two \Caf ions are entangled on an optical transition with a maximum target state fidelity of $96.9(3)\%$ at a magnetic field of 6 G. By introducing a waiting time between creation and the analysis of the Bell state we observe a Gaussian decay of Bell state fidelity ($\bullet$) (black solid line). Data for the first \unit[3]{ms} are displayed as inset. When the optical qubit is mapped to the hyperfine qubit subsequent to the gate operation we obtain a maximum Bell state fidelity of $96.7(3)\%$ and the lifetime of the entanglement increases to \unit[96(3)]{ms} ($\textcolor[rgb]{0.00,0.00,1.00}{\bigstar}$). Measurements taken at magnetic field of \unit[3.4]{G} give similar results (\textcolor[rgb]{0.98,0.00,0.00}{$\blacktriangle$}) for the coherence time. Here, the optical qubit $\qd \leftrightarrow D_{5/2}(F=4,m_F=2)$ in combination with the laser beam~2 was used for entangling the ions. Each data point represents $18.000$ to $24.000$ individual measurements.}
\end{figure}

By introducing a waiting time before the Bell state analysis we observe a Gaussian decay of the parity fringe contrast. From this model we extrapolate a Bell state fidelity of $75\%$ after \unit[3.43(5)]{ms} (see inset of Fig. \ref{fig:decay}) corresponding to a loss of $50\%$ of the coherence since the state populations hardly change. The magnetic field sensitivity of this optical qubit is \unit[350]{kHz/G} which is one cause of decoherence. However, in measurements where we probe a single \Caf \ ion on a first order field insensitive transition ($S_{1/2}(F=4,m_F=4) \leftrightarrow D_{5/2}(F=4,m_F=3)$ at $\unit[3.4]{G}$) we obtain single qubit coherence times of \unit[8.1(3)]{ms} and \unit[5.6(3)]{ms} depending on whether the laser was linked to the experiment with a \unit[2]{m} or a \unit[10]{m} long glass fiber. This leads us to the conclusion that the decoherence on the optical qubit is also caused by acoustical noise picked up by the \unit[2]{m} fiber cord used here and the finite linewidth ($\sim\unit[20]{Hz}$) of the laser.

Mapping the optical qubit to the hyperfine qubit is achieved by a $\pi$-pulse on the hyperfine qubit transition followed by a $\pi$-pulse on the optical qubit (map) which results in an average error rate of $0.2\%$. Detection of the hyperfine qubit states is done by shelving the population of \qd\ to the D$_{5/2}$ manifold by two consecutive $\pi$-pulses to different Zeeman states ($D_{5/2}(F=5,m_F=2)$ and $D_{5/2}(F=6,m_F=-2)$) followed by fluorescence detection. A Bell state fidelity measurement after the mapping yields a parity fringe contrast of $95.7\%$ - induced by varying the phase of a microwave $\pi/2$-pulse - corresponding to a target state fidelity of still $96.7\%$. By delaying the state analysis we again observe a Gaussian decay of the parity fringe pattern (Fig.~\ref{fig:decay}) corresponding to a $50\%$ loss of phase coherence on a timescale of \unit[96(3)]{ms} which means that we gain a factor $28$ in coherence time compared to the optical qubit.

For successive gate application the Bell state is mapped back to the optical qubit (map$^{-1}$) after a waiting time of 20~ms and the gate (MS~2) is applied a second time, disentangling the ions to \oquu \ in $90.4\%$ of the cases, indicating an error of $3.7\%$. Another subsequent application of the gate (MS~3) adds an error of $2.9\%$ leading to $\qdd-i\oquu$ with a fidelity of $87.8\%$.

The results with successive gates suggest that the Bell state fidelity depends slightly on whether we apply the gate to the input state \qdd \ or \oquu, an effect already observed in previous experiments. This observation can be explained by the different spectator states into which the population can leak by unwanted excitations. The effect is most prominent in a series of measurement where the magnetic field was set to \unit[3.4]{G}. For this field strength the Zeeman splitting of the ground state and the axial trap frequency coincide to within \unit[10]{kHz}. Using the transition
$\qd \leftrightarrow D_{5/2}(F=4,m_F=2)\equiv$~\oqub \ as optical qubit we adjust the polarization of beam~2 (see Fig.~\ref{fig:levelscheme}) to suppress the coupling strength to the nearest neighboring transitions. Note that in this geometry and polarization setting, both $\Delta m = \pm 2$ have optimal coupling, so a higher gate laser power is required as compared to \oqu\ . This leads to Bell state fidelities of up to $96.3 \%$ after a single gate operation applied to \qdd\ . The decoherence for these Bell states after mapping to the hyperfine qubit is also shown in Fig. \ref{fig:decay}. Applying the gate operation on the input state \oqubub \ the maximum obtained Bell state fidelity after a single gate operation is only $92\%$. Further measurements at a magnetic field of $\unit[1.36]{G}$ where we expected no coincidences of spectral components reveal a maximum Bell state fidelity of $95\%$. For these measurements, beam~1 was used again with $\qd \leftrightarrow \oqu$ as the optical qubit.


Comparing these results with the measurements taken with \Ca \ ions, we conclude that low errors of the M{\o}lmer-S{\o}rensen interaction on the optical qubit require that the qubit transition is well isolated from other spectral components such that no other transitions are erroneously excited. This view is supported by the observation that the gate mechanism favors a high magnetic field, where the transitions to neighboring lines are spectrally well separated. However, the sensitivity to magnetic field fluctuations of the hyperfine qubit increases for higher fields. Therefore, a particularly interesting regime for \Caf \ ions occurs at a magnetic field of \unit[150]{G} where nonlinearities in ground state Zeeman splitting lead to a first order magnetic field insensitive transition $S_{1/2}(F=4,m_F=0)\leftrightarrow S_{1/2}(F=3,m_F=1)$. This promises even longer coherence times as well as smaller errors for all spectrum dependent operations since the Zeeman splitting is huge compared to the measurements presented here.


In conclusion, we have entangled a pair of \Caf \ ions and demonstrated the combination of high fidelity entangling operations on one qubit system with the long storage time offered by another one. The coherence time exceeds the slowest operations by almost three orders of magnitude while the errors in the entanglement operation are only $2.3\%$ and the errors of all other operations are below $1\%$.

\begin{acknowledgments}
We gratefully acknowledge the support of the European network SCALA,
the Institut für Quanteninformationsverarbeitung and iARPA. R.~G. acknowledges funding by the Marie-Curie program of the
European Union (grant number PIEF-GA-2008-220105).
\end{acknowledgments}


\end{document}